\documentclass[aps,pra,twocolumn,superscriptaddress]{revtex4-1}
\usepackage{graphicx}% Include figure files
\usepackage{dcolumn}% Align table columns on decimal point
\usepackage{bm}% bold math
\usepackage{amsmath}
\usepackage{amssymb}
\usepackage{latexsym}
\usepackage{epsfig}
\usepackage{amsbsy}
\usepackage{array}
\usepackage{amssymb}
\usepackage{setspace}
\usepackage{bm}

\begin{document}
%%%%%%%%%%%
\title{Extraordinary behaviors in two-dimensional decoherent alternative quantum walk}
\date{\today}
\author{Tian Chen$^{}$}
%\email{chentian@bit.edu.cn}
\author{Xiangdong Zhang$^{}$}

\affiliation{$^{}$School of Physics, Beijing Institute of Technology, Beijing, 100081, China
}

\begin{abstract}
The quantum and classical behaviors of two-dimensional (2D) alternative quantum walk (AQW) in the presence of decoherence have been discussed in detail. For any kinds of decoherence, the analytic expressions for the moments of position distribution of AQW have been obtained. Taking the broken line noise and coin-decoherence as examples of decoherence, we find that when the decoherence only emerges in one direction, the anisotropic position distribution pattern appears, and not all the motions of quantum walkers exhibit the transition from quantum to classical behaviors. The correlations between the walkers and the coin in 2D AQW have been discussed. The anisotropic correlations between walkers and coin have been revealed in the presence of decoherence.

\end{abstract}

\maketitle

\section{Introduction}
The quantum walk (QW) has been widely employed as a useful tool to design quantum algorithms, quantum gates and quantum computation~\cite{Kempe_CP2003, Andraca_QIP2012, Farhi_PRA1998, Aharonov_PRA1993, Mackay_JPA2002, Inui_PRA2004, Childs_PRL2009, Childs_QIP2002, Shenvi_PRA2003, Childs_PRA2004, Ambainis_SIAM2004, Tulsi_PRA2008, Potocek_PRA2009, Abal_MSCS2010, Ghosh_PRA2014, Lyu_PRA2015, Chen_arXiv2015}. Due to the demand of searching from a large database, multi-dimensional quantum fast search algorithms based on quantum walks have drawn lots of attention~\cite{Shenvi_PRA2003, Childs_PRA2004, Ambainis_SIAM2004, Tulsi_PRA2008, Potocek_PRA2009, Abal_MSCS2010}. In the two-dimensional (2D) discrete time quantum walk (DTQW), by introducing the four-level Grover coin into the evolution, the quantum Grover search algorithm has been realized in the 2D position space~\cite{Ambainis_SIAM2004, Tulsi_PRA2008}. The quantum search in the higher-dimensional hypercube has also been discussed~\cite{Shenvi_PRA2003, Ambainis_SIAM2004, Potocek_PRA2009, Abal_MSCS2010}. When considering the real experimental implementation, the physical system will have an inevitable interaction with the surrounding environment. Many studies of DTQW report that due to the decoherence induced by the environment, the position distribution pattern of QW change to a binomial distribution that is similar to the distribution of the classical walk~\cite{Brun_PRL2003, Brun_PRA2003, Shapira_PRA2003, Romanelli_PhysA2005, Ermann_PRA2006, Kosik_PRA2006, Prokofec_PRA2006, Oliveira_PRA2006, Kendon_MSCS2007, Abal_PhysA2008, Romanelli_PRA2009, Annabestani_PRA2010, Liu_PRE2010, Liu_PRA2011, Romanelli_PhysA2011, Ampadu_CTP2012, Xue_PRA2013}. For the coherent QW, the variance of position distribution in the QW grows quadratically with time. While with the introduction of decoherence into the walk, the variance of the position will grows linearly with time, which is the characteristics of the classical walk. In some sense, the emergence of decoherence in the QW makes the original QW convert to a classical walk.

Recently, a 2D quantum walk with one two-level coin is presented~\cite{Franco_PRL2011, Franco_PRA2011, Roldan_PRA2013, Kollar_PRA2015, Franco_PRA2015}. In this alternative quantum walk (AQW), the two-level coin affects the walker moving in the $x$-direction at first, followed by the motion of walker along the $y$-direction. The position distribution pattern induced by the 2D DTQW with four-level Grover coin can be recovered with the AQW with only one two-level coin~\cite{Franco_PRL2011, Franco_PRA2011}. Due to the function of the coin as the register of the coherence and randomness in the aforementioned DTQW, when the searching space increases to $n$-dimension, we need a $2^n$-level coin to implement the search process. Such indispensable resource makes the quantum Grover search in high dimensions hardly realize in experiment. Thanks to the reduction of resource of AQW, this AQW is more feasible than the original 2D DTQW with a four-level coin in experimental realizations~\cite{Schreiber_Sci2012, Jeong_NC_2013}. The generalization of AQW to higher-dimensional position space has already been achieved~\cite{Roldan_PRA2013}. The effect of environment on the AQW needs to be counted in the applications of real-life quantum information. The decoherence including random phases, bit-flip noise, and phase flip noise in the coin space have already been discussed in the 2D AQW~\cite{Svozilik_PRA2012, Chandrashekar_arXiv2012, Chandrashekar_JPA2013}. The numerical results reveal that the classical behaviors will emerge in the 2D AQW with increasing the strength of the decoherence.

In this paper, we study the quantum and classical behaviors of the 2D AQW when the coin and the walker undergo any kinds of decoherence. By employing the method presented in Ref.~\cite{Brun_PRA2003, Annabestani_PRA2010}, we provide the analytic expressions for the first and second moments of position in the presence of any kinds of decoherence. We take the broken line noise model as an example of the coin-position decoherence of the AQW at first, then we consider a 2D AQW involving coin-decoherence that the coin is measured at each step with a certain probability. In our discussion, we assume that the decoherence emerges in the motion along the $x$-direction of 2D AQW. We study the position distribution of the 2D AQW, and the variance of the position distribution with the change of the strength of the decoherence. In our work, we find that, for different kinds of decoherence, the different quantum and classical behaviors in the 2D AQW emerge, and not both the motions along $x$ and $y$-directions of the AQW exhibit the transition from the quantum behaviors to classical behaviors. Different position distributions have been found between the 2D decoherent AQW and four-level coin Grover walk~\cite{Oliveira_PRA2006}. What's more, we study the classical and quantum correlations between the $x$ and $y$-directional walkers involving the coin-position decoherence and coin-decoherence. The anisotropic patterns for the correlations between quantum walkers and the coin have been found.

The organization of our work is as follows, the scheme of the 2D AQW incorporating any kinds of decoherence is introduced in Sec. \ref{II}. The first and second moments of position are addressed in analytic forms. Then in Sec. \ref{III}, we take the broken line noise model and coin decoherence model as examples. The variance of position and the position distribution of the AQW are presented and the anisotropic position distributions appear. The correlations between two directional walkers and the coin in 2D decoherent AQW are discussed in Sec.~\ref{IV}. A conclusion is given in Sec. \ref{V}.

\section{Model}\label{II}

In the 2D AQW, there exists two-directional walkers ($x$ and $y$) and one two-level coin ($|R\rangle$ and $|L\rangle$). The total Hilbert space for the walkers and coin are $\mathcal{H}_t=\mathcal{H}_x\bigotimes\mathcal{H}_y\bigotimes\mathcal{H}_c$. Here, $\mathcal{H}_x$  ($\mathcal{H}_y$) is an infinite dimensional Hilbert space, $\mathcal{H}_c$ is a two-level Hilbert space. The basis states of the space $\mathcal{H}_t$ are represented as $\{|x,y,c\rangle\}$, where $x$ and $y$ denote the position of the walkers along the $x$-direction and $y$-direction, respectively. The one step evolution of the 2D AQW consists of two conditional shift operations and two coin operations, $U_w=S_y(I\bigotimes C)S_x(I\bigotimes C)$. The coin operation $C$ is the Hadamard matrix, that is
\begin{equation}
C=H=\frac{1}{2}\left(\begin{array}{cc}
1 & 1\\
1 & -1\end{array}\right),
\end{equation}
followed by the conditional shift operation along the $x$-direction
\begin{equation}
S_x=\sum_{i,j\in\mathbb{Z}}|i-1,j,R\rangle\langle i,j,R|+\sum_{i,j\in\mathbb{Z}}|i+1,j,L\rangle\langle i,j,L|.
\end{equation}
After applying the operation $C$ on the coin space, the conditional shift operation along the $y$-direction is described as
\begin{equation}
S_y=\sum_{i,j\in\mathbb{Z}}|i,j-1,R\rangle\langle i,j,R|+\sum_{i,j\in\mathbb{Z}}|i,j+1,L\rangle\langle i,j,L|,
\end{equation}
where $\mathbb{Z}$ denotes the $x-y$ position space that is spanned by the Hilbert space $\mathcal{H}_x$ and $\mathcal{H}_y$. It has been verified that when the walkers start from the position $|0\rangle_x|0\rangle_y$, with an appropriate choice of the initial coin state, the position distribution of the 2D AQW at time $t$ is as same as that from the 2D DTQW with the four-level Grover coin~\cite{Franco_PRL2011, Franco_PRA2011}.

Due to the inevitable interaction with the surrounding environment, the evolution of the coherent AQW is affected by the noise. The one step evolution of system comprising the walkers and the coin can be written as the form of the Kraus operators~\cite{Nielsen_2000}
\begin{equation}
\rho(t+1)=\sum_{n=1}^m E_n\rho(t)E_n^\dagger.
\end{equation}
Here, the term $E_n$ is the Kraus operators containing the influence on the system from the environment. The complete relation for the Kraus operators is $\sum_{n=1}^m E_n^\dagger E_n=\mathbb{I}$. Here, in our discussion, the role of the system is represented by the walkers and coin of 2D AQW, the evolution of the system is affected by the noise from the surrounding environment. Considering that the evolution for the total system consisting of the system and environment is unitary, the total evolution for the total system can be addressed as~\cite{Brun_PRA2003, Annabestani_PRA2010}
\begin{equation}
U=|e_1\rangle\langle e_1|\bigotimes W_1+\cdots+|e_s\rangle\langle e_s|\bigotimes W_s,
\end{equation}
where for the environment state $|e_i\rangle$, the effect on the system is described by the operator $W_i$. Understanding the exact form of the environment is not required, and the effects of the environment on the system is contained in different operators $W_i$. With the introduction of the initial state for the environment
\begin{equation}
|env\rangle=\sqrt{f_1}|e_1\rangle+\sqrt{f_2}|e_2\rangle+\cdots+\sqrt{f_s}|e_s\rangle,
\end{equation}
the Kraus operators $E_n$ can be written as
\begin{equation}
E_n=\langle e_n|U|env\rangle=\sqrt{f_n}W_n.
\end{equation}
Here, the coefficient $f_n$ denotes the probability that the $n$th resource from the environment affects the system dynamics. Different forms of decoherence provide different expressions of $E_n$ and $W_n$. The one step evolution for the system is expressed as
\begin{equation}
\rho(t+1)=\sum_{n=1}^s E_n\rho(t)E_n^\dagger=\sum_{n=1}^s f_n W_n\rho(t)W_n^\dagger.
\end{equation}

The Fourier transform is applied to analyze the dynamics of the 2D AQW~\cite{Brun_PRA2003, Annabestani_PRA2010}. The transformations along $x$ and $y$-directions can be addressed as
\begin{equation}
\begin{split}
|x\rangle&=\int_{-\pi}^\pi\frac{dk}{2\pi}e^{-ikx}|k\rangle,\\
|y\rangle&=\int_{-\pi}^\pi\frac{dp}{2\pi}e^{-ipy}|p\rangle.
\end{split}
\end{equation}
Based on the equations above, we can formulate the expression for the element of position distribution as
\begin{equation}
\begin{split}
&\sum_{x,y}|x+l_1,y+l_2\rangle\langle x,y|\\&=\frac{1}{(2\pi)^2}\int_{-\pi}^{\pi}\int_{-\pi}^{\pi}dk dp e^{-il_1k-il_2p}|k,p\rangle\langle k,p|.
\end{split}
\end{equation}
The Kraus operator $E_n$ can be obtained in the form as
\begin{equation}
E_n=\frac{1}{(2\pi)^2}\int_{-\pi}^{\pi}\int_{-\pi}^{\pi}dk dp|k\rangle\langle k|\bigotimes|p\rangle\langle p|\bigotimes F_n(k,p).
\end{equation}
With the assumption that the walkers start from the position $(0,0)$ in the $x-y$ plane, the initial density matrix for the system is
\begin{equation}
\rho_0=\iiiint\frac{dk dk'}{4\pi^2}\frac{dp dp'}{4\pi^2}|k\rangle\langle k'|\bigotimes|p\rangle\langle p'|\bigotimes|\psi_0\rangle\langle\psi_0|,
\end{equation}
where the initial state of coin is represented by $|\psi_0\rangle$. After one step of evolution, the system can be formulated as
\begin{equation}\small
\begin{split}
\rho'&=\sum_{n=1}^mE_n\rho_0E_n^\dagger\\
&=\iiiint\frac{dk dk'}{4\pi^2}\frac{dp dp'}{4\pi^2}|k\rangle\langle k'|\bigotimes|p\rangle\langle p'|\bigotimes\mathcal{L}_{k,k',p,p'}|\psi_0\rangle\langle\psi_0|
\end{split}
\end{equation}
with $\mathcal{L}_{k,k',p,p'}\tilde{O}=\sum_nF_n(k,p)\tilde{O}F_n^\dagger(k',p')$. Thus, the system density matrix after $t$ step evolution can be obtained as
\begin{equation}\small
\begin{split}
&\rho(t)\\&=\iiiint\frac{dk dk'}{4\pi^2}\frac{dp dp'}{4\pi^2}|k\rangle\langle k'|\bigotimes|p\rangle\langle p'|\bigotimes\mathcal{L}^{t}_{k,k',p,p'}|\psi_0\rangle\langle\psi_0|.
\end{split}
\end{equation}
So at time $t$, the probability for the walkers occupying the position $(x,y)$ is
\begin{equation}\footnotesize
\begin{split}
&P(x,y,t)=\mathrm{Tr}_{x,y,c}[\rho(t)]\\&=\frac{1}{(2\pi)^4}\iiiint dk dk' dp dp' e^{-ix(k'-k)}e^{-iy(p'-p)}\mathrm{Tr}(\mathcal{L}^{t}_{k,k',p,p'}|\psi_0\rangle\langle\psi_0|).
\end{split}
\end{equation}
The $m$th moments of the probability distribution $\langle x^m\rangle$ and $\langle y^m\rangle$ for 2D AQW are defined as
\begin{equation}\label{16}\small
\begin{split}
&\langle x^m\rangle=\sum_{x,y}x^m P(x,y,t)\\
&=\frac{1}{(2\pi)^3}\sum_x x^m \iiint dk dk' dp e^{-ix(k'-k)}\mathrm{Tr}(\mathcal{L}^{t}_{k,k',p,p'}|\psi_0\rangle\langle\psi_0|),\\
&\langle y^m\rangle=\sum_{x,y}y^m P(x,y,t)\\
&=\frac{1}{(2\pi)^3}\sum_y y^m\iiint dk dk' dp e^{-iy(p'-p)}\mathrm{Tr}(\mathcal{L}^{t}_{k,k',p,p'}|\psi_0\rangle\langle\psi_0|).
\end{split}
\end{equation}
Based on the expressions for the $m$th moments of the position distribution, we can obtain the analytic forms for the first and second moments $\langle x\rangle$, $\langle y\rangle$, $\langle x^2\rangle$ and $\langle y^2\rangle$ in the presence of decoherence as
\begin{equation}\label{17}\footnotesize
\begin{split}
\langle x\rangle=&\frac{i}{(2\pi)^2}\iint dk dp\sum_{m=1}^t\mathrm{Tr}(\mathcal{K}_{k,p}\mathcal{L}_{k,p}^{m-1}|\psi_0\rangle\langle\psi_0|),\\
\langle y\rangle=&\frac{i}{(2\pi)^2}\iint dk dp\sum_{n=1}^t\mathrm{Tr}(\mathcal{P}_{k,p}\mathcal{L}_{k,p}^{n-1}|\psi_0\rangle\langle\psi_0|),\\
\langle x^2\rangle=&\frac{1}{(2\pi)^2}\iint dk dp\sum_{m=1}^t\sum_{m'=1}^{m-1}\{\mathrm{Tr}[\mathcal{K}_{k,p}\mathcal{L}_{k,p}^{m-m'-1}(\mathcal{K}_{k,p}^\dagger\mathcal{L}_{k,p}^{m'-1}|\psi_0\rangle\langle\psi_0|)]\\&+
\mathrm{Tr}[\mathcal{K}_{k,p}^\dagger\mathcal{L}_{k,p}^{m-m'-1}(\mathcal{K}_{k,p}\mathcal{L}_{k,p}^{m'-1}|\psi_0\rangle\langle\psi_0|)]\}\\&+\frac{1}{(2\pi)^2}\iint dk dp \sum_{m=1}^t\mathrm{Tr}
[\mathcal{T}_{k}(\mathcal{L}^{m-1}_{k,p}|\psi_0\rangle\langle\psi_0|)],\\
\langle y^2\rangle=&\frac{1}{(2\pi)^2}\iint dk dp\sum_{n=1}^t\sum_{n'=1}^{n-1}\{\mathrm{Tr}[\mathcal{P}_{k,p}\mathcal{L}_{k,p}^{n-n'-1}(\mathcal{P}_{k,p}^\dagger\mathcal{L}_{k,p}^{n'-1}|\psi_0\rangle\langle\psi_0|)]\\&+
\mathrm{Tr}[\mathcal{P}_{k,p}^\dagger\mathcal{L}_{k,p}^{n-n'-1}(\mathcal{P}_{k,p}\mathcal{L}_{k,p}^{n'-1}|\psi_0\rangle\langle\psi_0|)]\}\\&+\frac{1}{(2\pi)^2}\iint dk dp \sum_{n=1}^t\mathrm{Tr}
[\mathcal{T}_{p}(\mathcal{L}^{n-1}_{k,p}|\psi_0\rangle\langle\psi_0|)].
\end{split}
\end{equation}
Here, the superoperators $\mathcal{K}_{k,p}$, $\mathcal{K}^\dagger_{k,p}$, $\mathcal{T}_{k}$, $\mathcal{P}_{k,p}$, $\mathcal{P}^\dagger_{k,p}$ and $\mathcal{T}_{p}$ above are represented by the explicit expressions as $\mathcal{K}_{k,p}\tilde{O}=\sum_n \frac{\partial F_n}{\partial k}\tilde{O}F_n^\dagger$,
$\mathcal{K}^\dagger_{k,p}\tilde{O}=\sum_n F_n\tilde{O}\frac{\partial F_n^\dagger}{\partial k}$, $\mathcal{T}_{k}\tilde{O}=\sum_n\frac{\partial F_n}{\partial k}\tilde{O}\frac{\partial F_n^\dagger}{\partial k}$, $\mathcal{P}_{k,p}\tilde{O}=\sum_n \frac{\partial F_n}{\partial p}\tilde{O}F_n^\dagger$, $\mathcal{P}^\dagger_{k,p}\tilde{O}=\sum_n F_n\tilde{O}\frac{\partial F_n^\dagger}{\partial p}$ and $\mathcal{T}_{p}\tilde{O}=\sum_n\frac{\partial F_n}{\partial p}\tilde{O}\frac{\partial F_n^\dagger}{\partial p}$.

Based on the equations above, we have obtained the expressions for the moments of the position distribution (Eq.~\ref{16},~\ref{17}) for 2D AQW in the presence of any kings of decoherence already. In the following, we take the broken line noise and coin-decoherence as explicit forms of decoherence, and study the behaviors of 2D AQW under these two kinds of decoherence.

\section{Two kinds of decoherence for the two-dimensional AQW}\label{III}
In this section, at first, we take the broken line noise model as the example of the coin-position decoherence, then the coin-decoherence is introduced into the walk in which the coin is measured with a certain probability at each step of the AQW. We study the variances of position distribution and diffusion coefficients of these 2D decoherent AQW. The anisotropic position distribution patterns of these two decoherent AQW are presented later.

\subsection{The broken line noise model}
\begin{figure}[htbp]
\begin{center}
\includegraphics[width=0.5\textwidth]{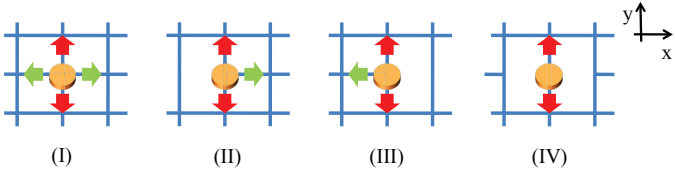}
\end{center}
\caption{\label{fig1} (Color online): A schematic representation for 2D AQW with broken line noise, the noise is applied only along the $x$-direction. Four possible cases of AQW are drawn as (I), no decoherence, with probability $(1-f)^2$, (II), the connection between the position and its left adjacent position is broken, with probability $f(1-f)$, (III), the connection between the position and its right adjacent position is broken, with probability $f(1-f)$, (IV), the connections between the position and its adjacent positions (left and right) are broken, with probability $f^2$. }
\end{figure}

\begin{figure}[htbp]
\begin{center}
\includegraphics[width=0.5\textwidth]{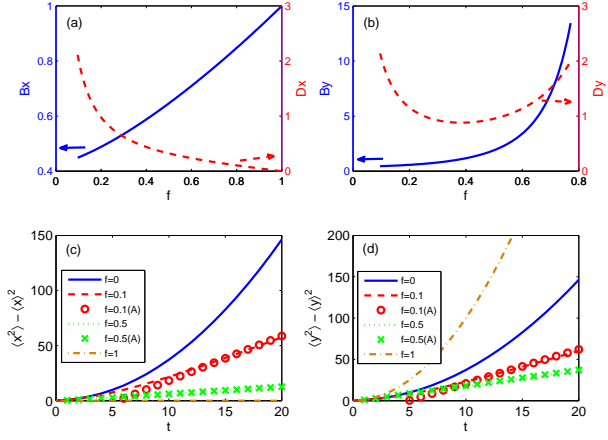}
\end{center}
\caption{\label{fig2} (Color online): The diffusion coefficients and the variances for 2D AQW with broken line noise, the noise is applied only along the $x$-direction. (a) and (b), the diffusion coefficients $D_x$ and $D_y$ with the probability $f$. (c) and (d), the variances of the position distribution of AQW with time, four different probabilities $f$ are presented. Blue solid, $f=0$; red dashed and red circle, $f=0.1$; green dotted and green cross, $f=0.5$; brown dotted dashed, $f=1$. The numerical results are expressed in the forms of solid, dashed, dotted, and dotted dashed lines. The analytic results from Eq.~\ref{32} and \ref{33} are shown in the forms of red circles and green crosses. In the numerical simulation, the initial state for the walkers and coin are, $|0\rangle_x|0\rangle_y\bigotimes(1/\sqrt{2}|R\rangle+i/\sqrt{2}|L\rangle)$. }
\end{figure}

In the quantum walk, the walker moves from one position to its adjacent positions controlled by the current state of coin. The walkers from different positions can interfere with each other when they meet at the same position simultaneously. That is the main reason for the different behaviors between the quantum and classical walk. For the case of quantum walk, the broken line noise model denotes one kind of decoherence that the connection between the position and its adjacent positions is broken with a certain probability~\cite{Romanelli_PhysA2005, Oliveira_PRA2006, Annabestani_PRA2010}. Here, we assume that the broken line noise appears only along the $x$-direction, and four possible evolutions of the 2D AQW involving decoherence are depicted in Fig.~\ref{fig1}.
In Fig.~\ref{fig1}, the walker moves in the $x$-direction at first, labeled by the green arrows along the horizontal direction; then the walker travels along the $y$-direction, labeled by the red arrows along the vertical direction. The 2D AQW without decoherence is addressed in Fig.~\ref{fig1} (I). The connection between the current position and its left adjacent position is broken, which is addressed in Fig.~\ref{fig1} (II). For Fig.~\ref{fig1} (III), the connection between the current position and its right adjacent position is cut off. In Fig.~\ref{fig1} (IV), both the right neighbour and left neighbour have no connection with the current position, and the motion along the $x$-direction is trapped in this case. Considering the description of the broken line noise above, the state for the environment can be addressed as
\begin{equation}
|env\rangle=(1-f)|e_1\rangle+\sqrt{f(1-f)}|e_2\rangle+\sqrt{f(1-f)}|e_3\rangle+f|e_4\rangle.
\end{equation}
We can obtain the expressions $F_n$, that are related with the corresponding Kraus operators $E_n$ as
\begin{subequations}\small
\begin{align}
F_1&=(1-f)\left(\begin{array}{cc}
\frac{1}{2}e^{-i(k+p)}+\frac{1}{2}e^{i(k-p)} & \frac{1}{2}e^{-i(k+p)}-\frac{1}{2}e^{i(k-p)}\\
\frac{1}{2}e^{-i(k-p)}-\frac{1}{2}e^{i(k+p)} & \frac{1}{2}e^{-i(k-p)}+\frac{1}{2}e^{i(k+p)}\\
\end{array}\right),\\
F_2&=\sqrt{f(1-f)}\left(\begin{array}{cc}
\frac{1}{2}e^{-i(k+p)}+\frac{1}{2}e^{-ip)} & \frac{1}{2}e^{-i(k+p)}-\frac{1}{2}e^{-ip}\\
\frac{1}{2}e^{-i(k-p)}+\frac{1}{2}e^{ip} & \frac{1}{2}e^{-i(k-p)}-\frac{1}{2}e^{ip}\\
\end{array}\right),\\
F_3&=\sqrt{(1-f)}\left(\begin{array}{cc}
\frac{1}{2}e^{-ip}-\frac{1}{2}e^{i(k-p)} & \frac{1}{2}e^{-ip}+\frac{1}{2}e^{i(k-p)}\\
-\frac{1}{2}e^{ip}+\frac{1}{2}e^{i(k+p)} & -\frac{1}{2}e^{ip}-\frac{1}{2}e^{i(k+p)}\\
\end{array}\right),\\
F_4&=f\left(\begin{array}{cc}
e^{-ip} & 0\\
0 & -e^{ip}\\
\end{array}\right).
\end{align}
\end{subequations}
To calculate the moments of the position ($\langle x\rangle$, $\langle y\rangle$, $\langle x^2\rangle$ and $\langle y^2\rangle$), we employ one representation that transforms one $2\times2$ matrix to one $4\times1$ column vector~\cite{Brun_PRA2003, Annabestani_PRA2010}, that is
\begin{equation}
\tilde{O}=r_0I+r_1\sigma_x+r_2\sigma_y+r_3\sigma_z=\left(\begin{array}{c}
r_0\\ r_1\\ r_2\\ r_3\end{array}\right).
\end{equation}
By using this representation, the superoperators $\mathcal{L}_{k,p}$, $\mathcal{K}_{k,p}$, $\mathcal{T}_{k}$, $\mathcal{P}_{k,p}$, and $\mathcal{T}_{p}$ of the 2D AQW involving the broken line noise can be obtained in the matrix form as
\begin{widetext}
\begin{equation}\scriptsize
\begin{split}
&\mathcal{L}_{k,p}\tilde{O}=\\&\left(\begin{array}{cccc}
1 & 0 & 0 & 0\\
0 & (1-2f)\cos2p & 2f(1-f)\sin k\cos2p-(1-f)^2\cos2k\sin2p+f^2\sin2p & (1-f)^2\sin2k\sin2p+2f(1-f)\cos k\cos2p\\
0 & (1-2f)\sin2p & (1-f)^2\cos2k\cos2p+2f(1-f)\sin k\sin2p-f^2\cos2p & 2f(1-f)\cos k\sin2p-(1-f)^2\cos2p\sin2k\\
0 & 0 & (1-f)^2\sin2k & (1-f)^2\cos2k+f^2\end{array}\right)\left(\begin{array}{c}
r_0\\ r_1\\ r_2\\ r_3\end{array}\right),
\end{split}
\end{equation}
\begin{equation}\scriptsize
\begin{split}
&\mathcal{K}_{k,p}\tilde{O}=\\&\left(\begin{array}{cccc}
0 & -i(1-f) & -if(1-f)\sin k & -if(1-f)\cos k\\
-i(1-f)\cos2p & 0 & f(1-f)\cos k\cos2p+(1-f)^2\sin2k\sin2p & (1-f)^2\cos2k\sin2p-f(1-f)\sin k\cos2p\\
-i(1-f)\sin2p & 0 & f(1-f)\cos k\sin2p-(1-f)^2\sin2k\cos2p & -f(1-f)\sin k\sin2p-(1-f)^2\cos2p\cos2k\\
0 & 0 & (1-f)^2\cos2k & -(1-f)^2\sin2k\end{array}\right)\left(\begin{array}{c}
r_0\\ r_1\\ r_2\\ r_3\end{array}\right),
\end{split}
\end{equation}
\begin{equation}\scriptsize
\begin{split}
\mathcal{T}_{k}\tilde{O}=\left(\begin{array}{cccc}
1-f & 0 & 0 & 0\\
0 & \cos2p(1-f) & (1-f)^2\cos2k\sin2p & -(1-f)^2\sin2k\sin2p\\
0 & \sin2p(1-f) & -(1-f)^2\cos2k\cos2p & (1-f)^2\cos2p\sin2k\\
0 & 0 & -(1-f)^2\sin2k & -(1-f)^2\cos2k\end{array}\right)\left(\begin{array}{c}
r_0\\ r_1\\ r_2\\ r_3\end{array}\right),
\end{split}
\end{equation}
\begin{equation}\scriptsize
\begin{split}
&\mathcal{P}_{k,p}\tilde{O}=\\&\left(\begin{array}{cccc}
0 & 0 & -i(1-f)^2\sin2k & -i(1-f)^2\cos2k-if^2\\
0 & (2f-1)\sin2p & -(1-f)^2\cos2k\cos2p-2f(1-f)\sin k\sin2p+f^2\cos2p & (1-f)^2\sin2k\cos2p-2f(1-f)\cos k\sin2p\\
0 & (1-2f)\cos2p & -(1-f)^2\cos2k\sin2p+2f(1-f)\sin k\cos2p+f^2\sin2p & (1-f)^2\sin2k\sin2p+2f(1-f)\cos k\cos2p\\
-i & 0 & 0 & 0\end{array}\right)\left(\begin{array}{c}
r_0\\ r_1\\ r_2\\ r_3\end{array}\right),
\end{split}
\end{equation}
\begin{equation}\scriptsize
\begin{split}
&\mathcal{T}_{p}\tilde{O}=\\&\left(\begin{array}{cccc}
1 & 0 & 0 & 0\\
0 & \cos2p(2f-1) & (1-f)^2\cos2k\sin2p-2f(1-f)\sin k\cos2p-f^2\sin2p & -(1-f)^2\sin2k\sin2p-2f(1-f)\cos k\cos2p\\
0 & \sin2p(2f-1) & -(1-f)^2\cos2k\cos2p-2f(1-f)\sin k\sin2p+f^2\cos2p & (1-f)^2\cos2p\sin2k-2f(1-f)\cos k\sin2p\\
0 & 0 & (1-f)^2\sin2k & f^2+(1-f)^2\cos2k\end{array}\right)\left(\begin{array}{c}
r_0\\ r_1\\ r_2\\ r_3\end{array}\right),
\end{split}
\end{equation}
\end{widetext}
where $\mathcal{K}^\dagger_{k,p}\tilde{O}=\mathcal{K}^*\tilde{O}$, $\mathcal{P}^\dagger_{k,p}\tilde{O}=\mathcal{P}^*\tilde{O}$, and the initial coin state $|\psi_0\rangle$ is set as
\begin{equation}
|\psi_0\rangle\langle\psi_0|=\left(\begin{array}{c}
r_0\\ r_1\\ r_2\\ r_3 \end{array}\right).
\end{equation}
Considering the expression of superoperator $\mathcal{L}$, we can verify that
\begin{equation}
\mathcal{L}_{k,p}^{m-1}|\psi_0\rangle\langle\psi_0|=\left(\begin{array}{c}
r_0 \\ r'_1 \\ r'_2 \\r'_3 \end{array}\right).
\end{equation}
The first row element $r_0$ keeps unchanged when any time of $\mathcal{L}$ is applied. When taking into account the trace operation, we obtain the results related to the operators $\mathcal{T}_k$ and $\mathcal{T}_p$
\begin{equation}
\begin{split}
\sum_{m=1}^t&\mathrm{Tr}[\mathcal{T}_k(\mathcal{L}_{k,p}^{m-1}|\psi_0\rangle\langle\psi_0|)]=2(1-f)t\cdot r_0,\\
\sum_{n=1}^t&\mathrm{Tr}[\mathcal{T}_p(\mathcal{L}_{k,p}^{n-1}|\psi_0\rangle\langle\psi_0|)]=2r_0t.
\end{split}
\end{equation}
Considering the expressions of the superoperators, the first row element $r_0$ of the $4\times1$ column vector have no contribution to the moments $\langle x^2\rangle$ and $\langle y^2\rangle$ (Eq.~\ref{17}), so we can omit the outcomes associated with $r_0$, and obtain the first term of the second moments $\langle x^2\rangle$, $\langle y^2\rangle$ as
\begin{widetext}
\begin{equation}%\footnotesize
\begin{split}
&\sum_{m=1}^t\sum_{m'=1}^{m-1}\{\mathrm{Tr}[\mathcal{K}_{k,p}\mathcal{L}_{k,p}^{m-m'-1}(\mathcal{K}_{k,p}^\dagger\mathcal{L}_{k,p}^{m'-1}|\psi_0\rangle\langle\psi_0|)]+
\mathrm{Tr}[\mathcal{K}_{k,p}^\dagger\mathcal{L}_{k,p}^{m-m'-1}(\mathcal{K}_{k,p}\mathcal{L}_{k,p}^{m'-1}|\psi_0\rangle\langle\psi_0|)]\}\\&
=4(1-f,f(1-f)\sin k, f(1-f)\cos k)\sum_{m=1}^t\sum_{m'=1}^{m-1}\mathcal{M}_{k,p}^{m-m'-1}\left(\begin{array}{c}
(1-f)\cos2p\cdot r_0\\ (1-f)\sin2p\cdot r_0\\ 0 \end{array}\right)\\&
=2(1-f,f(1-f)\sin k, f(1-f)\cos k)(I-\mathcal{M}_{k,p})^{-1}\{t-\frac{\mathcal{M}_{k,p}}{I-\mathcal{M}_{k,p}}\}\left(\begin{array}{c}
(1-f)\cos2p\\ (1-f)\sin2p\\ 0 \end{array}\right),
\end{split}
\end{equation}
\begin{equation}%\footnotesize
\begin{split}
&\sum_{n=1}^t\sum_{n'=1}^{n-1}\{\mathrm{Tr}[\mathcal{P}_{k,p}\mathcal{L}_{k,p}^{n-n'-1}(\mathcal{P}_{k,p}^\dagger\mathcal{L}_{k,p}^{m'-1}|\psi_0\rangle\langle\psi_0|)]+
\mathrm{Tr}[\mathcal{P}_{k,p}^\dagger\mathcal{L}_{k,p}^{n-n'-1}(\mathcal{P}_{k,p}\mathcal{L}_{k,p}^{n'-1}|\psi_0\rangle\langle\psi_0|)]\}\\&
=4(0,(1-f)^2\sin2k, (1-f)^2\cos2k+f^2)\sum_{n=1}^t\sum_{n'=1}^{n-1}\mathcal{M}_{k,p}^{n-n'-1}\left(\begin{array}{c}
0\\ 0\\ r_0 \end{array}\right)\\&
=2(0,(1-f)^2\sin2k, (1-f)^2\cos2k+f^2)(I-\mathcal{M}_{k,p})^{-1}\{t-\frac{\mathcal{M}_{k,p}}{I-\mathcal{M}_{k,p}}\}\left(\begin{array}{c}
0\\ 0\\ 1 \end{array}\right),
\end{split}
\end{equation}
\end{widetext}
where $r_0$ is chosen as $r_0=1/2$ for the normalization of the initial state $|\psi_0\rangle$. The term $\mathcal{M}_{k,p}$ is a $3\times3$ matrix,
\begin{widetext}
\begin{equation}\scriptsize
\begin{split}
&\mathcal{M}_{k,p}\tilde{O}=\\&\left(\begin{array}{ccc}
(1-2f)\cos2p & 2f(1-f)\sin k\cos2p-(1-f)^2\cos2k\sin2p+f^2\sin2p & (1-f)^2\sin2k\sin2p+2f(1-f)\cos k\cos2p\\
(1-2f)\sin2p & (1-f)^2\cos2k\cos2p+2f(1-f)\sin k\sin2p-f^2\cos2p & 2f(1-f)\cos k\sin2p-(1-f)^2\cos2p\sin2k\\
0 & (1-f)^2\sin2k & (1-f)^2\cos2k+f^2\end{array}\right)\left(\begin{array}{c}
r_1\\ r_2\\ r_3\end{array}\right).
\end{split}
\end{equation}
\end{widetext}
Based on the equations addressed above, the first moments of the position $\langle x\rangle$ and $\langle y\rangle$ for the 2D AQW with broken line noise are presented as
\begin{equation}\footnotesize\label{32}
\begin{split}
&\langle x\rangle\\&=\frac{i}{2\pi^2}\iint dk dp (-i)\left(\begin{array}{c}
1-f \\ f(1-f)\sin k\\ f(1-f)\cos k\end{array}\right)^T[\sum_{m=1}^t\mathcal{M}_{k,p}^{m-1}]\left(\begin{array}{c}
r_1\\ r_2\\ r_3 \end{array}\right)\\
&=\frac{1}{2\pi^2}\iint dk dp \left(\begin{array}{c}
1-f \\ f(1-f)\sin k\\ f(1-f)\cos k\end{array}\right)^T(I-\mathcal{M}_{k,p})^{-1}\left(\begin{array}{c}
r_1\\ r_2\\ r_3 \end{array}\right),\\
&\langle y\rangle\\&=\frac{i}{2\pi^2}\iint dk dp (-i)\left(\begin{array}{c}
0\\ (1-f)^2\sin2k \\ (1-f)^2\cos2k+f^2 \end{array}\right)^T[\sum_{n=1}^t\mathcal{M}_{k,p}^{n-1}]\left(\begin{array}{c}
r_1\\ r_2\\ r_3 \end{array}\right)\\
&=\frac{1}{2\pi^2}\iint dk dp \left(\begin{array}{c}
0\\ (1-f)^2\sin2k \\ (1-f)^2\cos2k+f^2 \end{array}\right)^T(I-\mathcal{M}_{k,p})^{-1}\left(\begin{array}{c}
r_1\\ r_2\\ r_3 \end{array}\right),
\end{split}
\end{equation}
where the superscript $T$ stands for the transpose on that matrix, and the second moments of the position $\langle x^2\rangle$ and $\langle y^2\rangle$ for the 2D AQW with broken line noise are addressed as
\begin{equation}\footnotesize\label{33}
\begin{split}
&\langle x^2\rangle\\&=\frac{1}{2\pi^2}\iint dk dp\{ \left(\begin{array}{c}
1-f\\ f(1-f)\sin k \\ f(1-f)\cos k \end{array}\right)^T(I-\mathcal{M}_{k,p})^{-1}[t-\frac{\mathcal{M}_{k,p}}{I-\mathcal{M}_{k,p}}]\\&\cdot\left(\begin{array}{c}
(1-f)\cos2p\\ (1-f)\sin2p\\ 0 \end{array}\right)\}+\frac{1}{2\pi^2}\iint dk dp \frac{1}{2}(1-f)t.\\
&\langle y^2\rangle\\&=\frac{1}{2\pi^2}\iint dk dp\{ \left(\begin{array}{c}
0 \\ (1-f)^2\sin2k\\ (1-f)^2\cos2k+f^2\end{array}\right)^T(I-\mathcal{M}_{k,p})^{-1}\\&\cdot[t-\frac{\mathcal{M}_{k,p}}{I-\mathcal{M}_{k,p}}]\left(\begin{array}{c}
0\\ 0\\ 1 \end{array}\right)\}+\frac{1}{2\pi^2}\iint dk dp \frac{1}{2}t.
\end{split}
\end{equation}
To illustrate the transition from quantum to classical behaviors in the 2D AQW involving the broken line noise, we calculate the diffusion coefficients $D_x$ and $D_y$ as
\begin{equation}\scriptsize\label{34}
\begin{split}
D_x=&\frac{1}{2}\lim_{t\rightarrow\infty}\frac{\partial\sigma_x^2}{\partial t}=\frac{1}{2}\lim_{t\rightarrow\infty}\frac{\partial(\langle x^2\rangle-\langle x\rangle^2)}{\partial t}\\
=&\frac{1}{2}\{\frac{1}{2\pi^2}\iint dk dp \left(\begin{array}{c}
1-f\\f(1-f)\sin k\\ f(1-f)\cos k \end{array}\right)^T
(I-\mathcal{M}_{k,p})^{-1}\left(\begin{array}{c}
(1-f)\cos2p\\ (1-f)\sin2p\\ 0 \end{array}\right)\\&+(1-f)\}\\
=&\frac{1-f}{2f}[\frac{1}{2\pi^2}\iint dk dp \frac{(1-f)}{2}R_x(f,k,p)+f]=\frac{1-f}{2f}B_x(f),\\
D_y=&\frac{1}{2}\lim_{t\rightarrow\infty}\frac{\partial\sigma_y^2}{\partial t}=\frac{1}{2}\lim_{t\rightarrow\infty}\frac{\partial(\langle y^2\rangle-\langle y\rangle^2)}{\partial t}\\
=&\frac{1}{2}\{\frac{1}{2\pi^2}\iint dk dp \left(\begin{array}{c}
0\\ (1-f)^2\sin2k\\ (1-f)^2\cos2k+f^2\end{array}\right)^T(I-\mathcal{M}_{k,p})^{-1}\left(\begin{array}{c}
0\\ 0\\ 1 \end{array}\right)+1\}\\
=&\frac{1-f}{2f}[\frac{1}{2\pi^2}\iint dk dp R_y(f,k,p)+\frac{f}{1-f}]=\frac{1-f}{2f}B_y(f).
\end{split}
\end{equation}
The expressions of $R_x$ and $R_y$ can be obtained analytically. Due to the lengthy expressions, we do not present the explicit forms for them here. The change of diffusion coefficients $D_x$ and $D_y$ with the probability $f$ (red dashed lines) are shown in Fig.~\ref{fig2} (a) and (b). Besides, the terms $B_x$ and $B_y$ are addressed in blue solid lines. We can find that when there is no decoherence in the AQW, that is $f=0$, both diffusion coefficients $D_x$ and $D_y$ are not linear dependent on the time. With the increase of the probability $f$, the variance along the $x$-direction and $y$-direction both change to grow linearly with time, that means when the connection has a certain probability to be broken in the adjacent positions of AQW, both motions along the $x$-direction and $y$-direction exhibit classical behaviors. When the probability $f$ approaches $1$, the diffusion coefficient along the $x$-direction $D_x$ approaches zero, but the diffusion coefficient along the $y$-direction $D_y$ approaches infinity. That is when the probability $f$ becomes larger enough, the only remaining case (IV in Fig.~\ref{fig1}) of motion traps the walker along the $x$-direction with higher probability, and the other three cases of motions (I, II, III in Fig.~\ref{fig1}) are suppressed. At this time, the motion along the $x$-direction has little effect on the motion along the $y$-direction. So as shown in Fig.~\ref{fig2} (a) and (b), the diffusion coefficient of position in the $x$-direction $D_x$ is close to zero, while in the $y$-direction $D_y$ goes back to infinity, and the motion of the $y$-direction exhibits the quantum behavior.
In Fig.~\ref{fig2} (c) and (d), we numerically calculate the variances of the position distribution along $x$ and $y$-directiona with time, and compare them with the obtained analytic expressions for the variances in the long time limit (Eq.~\ref{32},~\ref{33}, and~\ref{34}). In these two figures, four different probabilities $f$ are chosen for comparison. The blue solid, red dashed, green dotted and brown dotted dashed lines correspond to the variance obtained numerically with probability $f$ chosen as $0$, $0.1$, $0.5$ and $1$, respectively. The analytic results from Eq.~\ref{32} and \ref{33} are addressed as the red circle and green cross in two figures. In our numerical simulation, we take the initial coin state as $|\psi_0\rangle=1/\sqrt{2}|R\rangle+i/\sqrt{2}|L\rangle$. From the figures, we find that, when the probability $f$ is zero (blue solid lines), both the motions along the $x$-direction and $y$-direction display quantum properties. As the probability $f$ becomes larger (red and green lines), both the variances of motion ($\sigma_x^2$ and $\sigma_y^2$) change to grow linearly with time and the classical behaviors emerge in both $x$ and $y$ directions of AQW. While, when the probability $f$ is $1$ (brown lines), for the walker in the $x$-direction is trapped at current positions, the variance of motion along the $x$-direction is zero, but the behavior of motion in the $y$-direction return back to the quantum region. As stated above, our numerical results coincide with the analytic results obtained in Eq.~\ref{32} and \ref{33}. For different broken probabilities, the time change of variances for the $x$-direction and $y$-direction in Fig.~\ref{fig2} (c) and (d) reflect the transitions of motion from quantum to classical behaviors, which have been presented in Fig.~\ref{fig2} (a) and (b).

\subsection{The coin-decoherence model}
\begin{figure}[htbp]
\begin{center}
\includegraphics[width=0.5\textwidth]{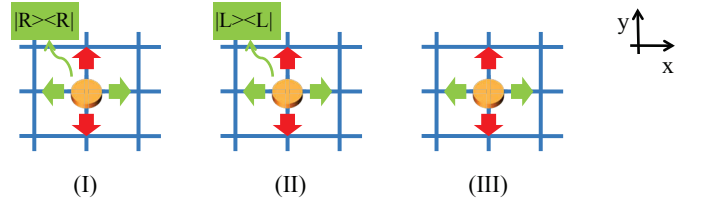}
\end{center}
\caption{\label{fig3} (Color online): A schematic representation for 2D AQW with coin-decoherence, the noise is applied before each step of walking. Three possible cases of AQW are drawn as (I) and (II), the coin is measured with probability $f$ at each step evolution, (III), no decoherence in the walk, with probability $1-f$.}
\end{figure}

In this section, we study the case that the decoherence only appear in the coin space of the 2D AQW. At each step evolution of 2D AQW, the coin is measured with a certain probability, which makes the coin state decoherence. In our discussion, we assume that such projective measurements emerge before each step of walking~\cite{Brun_PRA2003}. The schematic representation of possible evolutions of AQW involving coin-decoherence is depicted in Fig.~\ref{fig3},
In Fig.~\ref{fig3}, the walker moves in the $x$-direction at first, labeled by the green arrows along the horizontal direction; then the walker travels along the $y$-direction, labeled by the red arrows along the vertical direction. Due to the measurement, the coin is projected onto the states $|R\rangle\langle R|$ and $|L\rangle\langle L|$, see Fig.~\ref{fig3} (I) and (II), respectively. The 2D AQW without decoherence is addressed in Fig.~\ref{fig3} (III). When taking into this kind of decoherence, the expressions $F_n$ related to the Kraus operators $E_n$ for the 2D AQW involving coin-decoherence can be represented as
\begin{equation}
\begin{split}
F_1&=\sqrt{f}\left(\begin{array}{cc}
\frac{1}{2}e^{-i(k+p)}+\frac{1}{2}e^{i(k-p)} & 0\\
\frac{1}{2}e^{-i(k-p)}-\frac{1}{2}e^{i(k+p)} & 0 \end{array}\right),\\
F_2&=\sqrt{f}\left(\begin{array}{cc}
0& \frac{1}{2}e^{-i(k+p)}-\frac{1}{2}e^{i(k-p)}\\
0& \frac{1}{2}e^{-i(k-p)}+\frac{1}{2}e^{i(k+p)} \end{array}\right),\\
F_3&=\sqrt{1-f}\left(\begin{array}{cc}
\frac{1}{2}e^{-i(k+p)}+\frac{1}{2}e^{i(k-p)} & \frac{1}{2}e^{-i(k+p)}-\frac{1}{2}e^{i(k-p)}\\
\frac{1}{2}e^{-i(k-p)}-\frac{1}{2}e^{i(k+p)} & \frac{1}{2}e^{-i(k-p)}+\frac{1}{2}e^{i(k+p)} \end{array}\right).
\end{split}
\end{equation}
The complete relation is satisfied with $\sum_nF_n^\dagger F_n=I$. Using the similar technique mentioned in the section of broken line noise model, we take a $4\times1$ column vector to represent the $2\times2$ matrix, and the expressions for the superoperators $\mathcal{L}_{k,p}$, $\mathcal{K}_{k,p}$, $\mathcal{P}_{k,p}$, $\mathcal{T}_k$ and $\mathcal{T}_p$ are
\begin{widetext}
\begin{equation}\footnotesize
\begin{split}
&\mathcal{L}_{k,p}\tilde{O}=\left(\begin{array}{cccc}
1 & 0 & 0 & 0\\
0 & (1-f)\cos2p & -(1-f)\cos2k\sin2p & \sin2k\sin2p\\
0 & (1-f)\sin2p & (1-f)\cos2k\cos2p & -\sin2k\cos2p\\
0 & 0 & (1-f)\sin2k & \cos2k\end{array}\right)\left(\begin{array}{c}
r_0\\ r_1\\ r_2\\ r_3\end{array}\right),
\end{split}
\end{equation}
\begin{equation}\footnotesize
\begin{split}
&\mathcal{K}_{k,p}\tilde{O}=\left(\begin{array}{cccc}
0 & i(f-1) & 0 & 0\\
-i\cos2p & 0 & (1-f)\sin2k\sin2p & \cos2k\sin2p\\
-i\sin2p & 0 & -(1-f)\sin2k\cos2p & -\cos2k\cos2p\\
0 & 0 & (1-f)\cos2k & -\sin2k\end{array}\right)\left(\begin{array}{c}
r_0\\ r_1\\ r_2\\ r_3\end{array}\right),
\end{split}
\end{equation}
\begin{equation}\footnotesize
\begin{split}
&\mathcal{P}_{k,p}\tilde{O}=\left(\begin{array}{cccc}
0 & 0 & i(f-1)\sin2k & -i\cos2k\\
0 & -(1-f)\sin2p & -(1-f)\cos2k\cos2p & \sin2k\cos2p\\
0 & (1-f)\cos2p & -(1-f)\cos2k\sin2p & \sin2k\sin2p\\
-i & 0 & 0 & 0\end{array}\right)\left(\begin{array}{c}
r_0\\ r_1\\ r_2\\ r_3\end{array}\right),
\end{split}
\end{equation}
\begin{equation}\footnotesize
\begin{split}
&\mathcal{T}_{k}\tilde{O}=\left(\begin{array}{cccc}
1 & 0 & 0 & 0\\
0 & (1-f)\cos2p & (1-f)\cos2k\sin2p & -\sin2k\sin2p\\
0 & (1-f)\sin2p & -(1-f)\cos2k\cos2p & \sin2k\cos2p\\
0 & 0 & -(1-f)\sin2k & -\cos2k\end{array}\right)\left(\begin{array}{c}
r_0\\ r_1\\ r_2\\ r_3\end{array}\right),
\end{split}
\end{equation}
\begin{equation}\footnotesize
\begin{split}
&\mathcal{T}_{p}\tilde{O}=\left(\begin{array}{cccc}
1 & 0 & 0 & 0\\
0 & -(1-f)\cos2p & (1-f)\cos2k\sin2p & -\sin2k\sin2p\\
0 & -(1-f)\sin2p & -(1-f)\cos2k\cos2p & \sin2k\cos2p\\
0 & 0 & (1-f)\sin2k & \cos2k\end{array}\right)\left(\begin{array}{c}
r_0\\ r_1\\ r_2\\ r_3\end{array}\right),
\end{split}
\end{equation}
\end{widetext}
where $\mathcal{K}^\dagger_{k,p}\tilde{O}=\mathcal{K}^*\tilde{O}$, $\mathcal{P}^\dagger_{k,p}\tilde{O}=\mathcal{P}^*\tilde{O}$. With these superoperators, the first moments of position distribution $\langle x\rangle$ and $\langle y\rangle$ in 2D AQW with coin-decoherence are obtained as
\begin{equation}\footnotesize
\begin{split}
\langle x\rangle&=\frac{i}{2\pi^2}\iint dk dp (-i)(1-f, 0, 0)[\sum_{m=1}^t\mathcal{M}_{k,p}^{m-1}]\left(\begin{array}{c}
r_1\\ r_2\\ r_3 \end{array}\right)\\
&=\frac{1}{2\pi^2}\iint dk dp (1-f, 0, 0)(I-\mathcal{M}_{k,p})^{-1}\left(\begin{array}{c}
r_1\\ r_2\\ r_3 \end{array}\right),\\
\langle y\rangle&=\frac{i}{2\pi^2}\iint dk dp (-i)(0, (1-f)\sin2k, \cos2k)[\sum_{n=1}^t\mathcal{M}_{k,p}^{n-1}]\left(\begin{array}{c}
r_1\\ r_2\\ r_3 \end{array}\right)\\
&=\frac{1}{2\pi^2}\iint dk dp (0, (1-f)\sin2k, \cos2k)(I-\mathcal{M}_{k,p})^{-1}\left(\begin{array}{c}
r_1\\ r_2\\ r_3 \end{array}\right).
\end{split}
\end{equation}
The second moments of position distribution $\langle x^2\rangle$ and $\langle y^2\rangle$ in 2D AQW with coin-decoherence are
\begin{equation}\scriptsize
\begin{split}
&\langle x^2\rangle\\=&\frac{1}{2\pi^2}\iint dk dp \left(\begin{array}{c}
1-f\\ 0\\ 0\end{array}\right)^T(I-\mathcal{M}_{k,p})^{-1}\{t-\frac{\mathcal{M}_{k,p}}{I-\mathcal{M}_{k,p}}\}\left(\begin{array}{c}
\cos2p\\ \sin2p\\ 0 \end{array}\right)\\&+\frac{1}{2\pi^2}\iint dk dp \frac{1}{2}\cdot t,\\
&\langle y^2\rangle\\=&\frac{1}{2\pi^2}\iint dk dp \left(\begin{array}{c}
0\\ (1-f)\sin2k\\ \cos2k\end{array}\right)^T(I-\mathcal{M}_{k,p})^{-1}\{t-\frac{\mathcal{M}_{k,p}}{I-\mathcal{M}_{k,p}}\}\left(\begin{array}{c}
0\\ 0\\ 1 \end{array}\right)\\&+\frac{1}{2\pi^2}\iint dk dp \frac{1}{2}\cdot t.
\end{split}
\end{equation}
The diffusion coefficients $D_x$ and $D_y$ are addressed as
\begin{equation}\footnotesize
\begin{split}
D_x=&\frac{1}{2}\lim_{t\rightarrow\infty}\frac{\partial\sigma_x^2}{\partial t}=\frac{1}{2}\lim_{t\rightarrow\infty}\frac{\partial(\langle x^2\rangle-\langle x\rangle^2)}{\partial t}\\
=&\frac{1}{2}\{\frac{1}{2\pi^2}\iint dk dp \left(\begin{array}{c}
1-f\\ 0\\ 0\end{array}\right)^T(I-\mathcal{M}_{k,p})^{-1}\left(\begin{array}{c}
\cos2p\\ \sin2p\\ 0 \end{array}\right)\\&+\frac{1}{2\pi^2}\iint dk dp \frac{1}{2}\}\\
=&\frac{1-f}{2f}[\frac{1}{2\pi^2}\iint dk dp R_x(f,k,p)+\frac{f}{1-f}]=\frac{1-f}{2f}B_x(f),\\
D_y=&\frac{1}{2}\lim_{t\rightarrow\infty}\frac{\partial\sigma_y^2}{\partial t}=\frac{1}{2}\lim_{t\rightarrow\infty}\frac{\partial(\langle y^2\rangle-\langle y\rangle^2)}{\partial t}\\
=&\frac{1}{2}\{\frac{1}{2\pi^2}\iint dk dp \left(\begin{array}{c}
0\\ (1-f)\sin2k\\ \cos2k\end{array}\right)^T(I-\mathcal{M}_{k,p})^{-1}\left(\begin{array}{c}
0\\ 0\\ 1 \end{array}\right)\\&+\frac{1}{2\pi^2}\iint dk dp \frac{1}{2}\}\\
=&\frac{1-f}{2f}[\frac{1}{2\pi^2}\iint dk dp R_y(f,k,p)+\frac{f}{1-f}]=\frac{1-f}{2f}B_y(f).
\end{split}
\end{equation}
Here, the term $R_x=\frac{1-f+\cos2p}{2-f}$, and $R_y=\frac{-2(1-f)\cos2p\cot^2k+[(1-f)^2+\cos2k]\csc^2k}{2(2-3f+f^2)}$.
\begin{figure}[htbp]
\begin{center}
\includegraphics[width=0.5\textwidth]{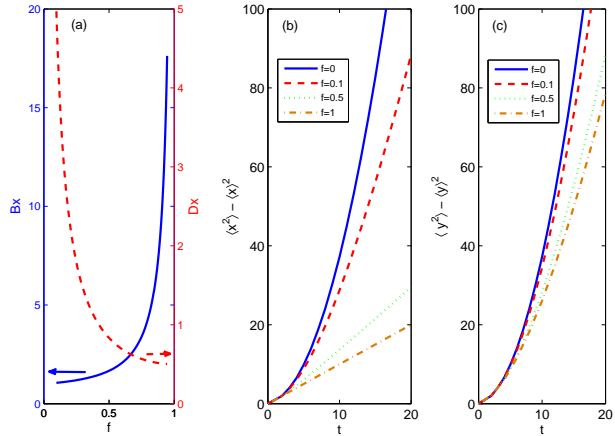}
\end{center}
\caption{\label{fig4} (Color online): The diffusion coefficient and the variances for 2D AQW with coin-decoherence, the noise is applied only along the $x$-direction. (a), the diffusion coefficient $D_x$ (red dashed line) and $B_x$ (blue solid line) with the probability $f$. (b) and (c), the variances of the position distribution of AQW with time, four different probabilities $f$ are chosen. Blue solid, $f=0$; red dashed, $f=0.1$; green dotted, $f=0.5$; brown dotted dashed, $f=1$. The initial state for the walkers and coin are, $|0\rangle_x|0\rangle_y\bigotimes(1/\sqrt{2}|R\rangle+i/\sqrt{2}|L\rangle)$. }
\end{figure}
We find that, the function $R_y(f,k,p)$ approaches infinity at some values of $k$ and $p$ for any values of $f$, which makes the amplitude of diffusion coefficients $D_y$ approach infinity. It means the motion along the $y$-direction always exhibit the quantum behavior, whatever the probability $f$ is taken. The diffusion coefficient $D_x$ with the change of the probability $f$ is presented in Fig.~\ref{fig4} (a). The diffusion coefficient $D_x$ is addressed in the red dashed line, and the term $B_x$ is presented in the blue solid line. For the walker travels along the $x$-direction, with the increase of the probability $f$, the effect of decoherence on the system becomes stronger, and the variance along the $x$-direction reveals the linear time-dependence. When the probability $f$ approaches $1$, the motion along the $x$-direction exhibits the classical behavior with the diffusion coefficient $D_x$ equals to $1/2$. The variances of position distribution along $x$ and $y$-directions reveal different dependence on the decoherent strength $f$. In our study, the coin-decoherence appears before each step of walking, the coherence of the coin is lost and the motion along the $x$-direction is affected, but the motion along the $y$-direction feels little effect and always displays the quantum behavior with the change of probability $f$. In Fig.~\ref{fig4} (b) and (c), we numerically calculate the variance of the position distribution along $x$ and $y$-directions of the 2D AQW involving coin-decoherence with time. Four different probabilities $f$ are chosen for comparison, The blue solid, red dashed, green dotted and brown dotted dashed lines correspond to the probability $f$ chosen as $f=0$, $f=0.1$, $f=0.5$ and $f=1$, respectively. From these two figures, with the increase of the probability $f$, the decoherence becomes stronger ($f=0.5$ and $f=1$), we can find that the variance along the $x$-direction changes to grow linearly with time. The motion in this direction displays the classical behaviors. In comparison, the motion along the $y$-direction always exhibits quantum behaviors for any values of probability $f$. Our numerical results in Fig.~\ref{fig4} (b) and (c) coincide with the analytic statements in Fig.~\ref{fig4} (a). When comparing the 2D AQW involving the broken line noise and coin-decoherence, different behaviors for variances of position distributions have been uncovered. In the former case, the decoherence affects both the coin and position, when the decoherence on the $x$-direction happens, the motion along the $x$-direction has the influence on the state of the coin, which in sequence affects the motion along the $y$-direction, while, in the latter case, the decoherence affects only the coin, the motion along the $y$-direction is influenced little by the coin-decoherence on the $x$-direction of AQW. So the motion along the $y$-direction keeps its quantum properties with the change of decoherence strength.

\begin{figure}[htbp]
\begin{center}
\includegraphics[width=0.5\textwidth]{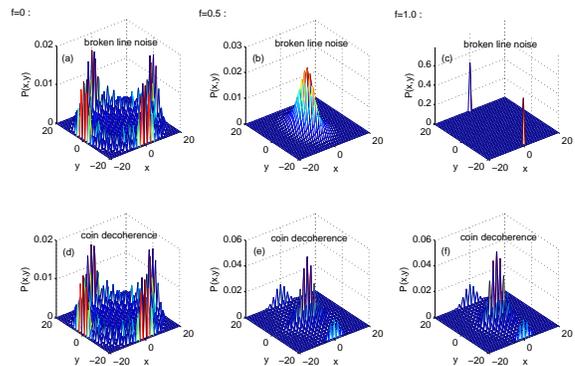}
\end{center}
\caption{\label{fig5} (Color online): Probability distribution on the position space $x-y$ of 2D AQW involving two kinds of decoherence at time step $t=20$. (a), (b) and (c), the broken line noise is introduced in the $x$-direction; (d), (e) and (f), the coin-decoherence is applied on the coin before the walker travels along the $x$-direction. Three different probabilities $f$ are chosen, leftmost ((a) and (d)), $f=0$; middle ((b) and (e)), $f=0.5$; rightmost ((c) and (f)), $f=1$. The initial state for the walkers and coin are, $|0\rangle_x|0\rangle_y\bigotimes(1/\sqrt{2}|R\rangle+i/\sqrt{2}|L\rangle)$.}
\end{figure}

To illustrate effects of different kinds of decoherence and reveal the anisotropic behaviors in the position space, we present the probability distributions on the $x-y$ position space of 2D AQW with different decoherence strengths in Fig.~\ref{fig5}.
The three figures above describe the position distribution of the 2D AQW with the broken line noise, and the three figures below denote the case of 2D decoherent AQW with the coin-decoherence. The time steps of all figures are $20$. Three different probabilities $f$ are chosen, two figures in the leftmost, $f=0$; two figures in the middle, $f=0.5$; two figures in the rightmost, $f=1$. Compared with the no-decoherence AQW ($f=0$, in Fig.~\ref{fig5}, that is (a) and (d)), when the decoherence is introduced (Fig.~\ref{fig5} (b), (c), (e) and (f)), the interference pattern of the probability distribution in the AQW changes, and the anisotropic distribution on the $x-y$ position space of 2D AQW can be found. With the increase of probability $f$ (Fig.~\ref{fig5} (b) and (e)), the position distribution becomes the binomial distribution, not the complex, oscillatory form as described in the coherent QW. When the probability $f$ is $1$, for the walk is affected by the broken line noise (Fig.~\ref{fig5} (c)), the position distribution along the $x$-direction is trapped, which coincides with the value of zero for the variance along the $x$-direction (Fig.~\ref{fig2}). The position distribution along the $y$-direction still spreads with time. Due to one coin has been tossed twice in one step evolution of AQW, this position distribution is different from the case of four-level coin decoherent Grover walk~\cite{Oliveira_PRA2006}. For the walk with coin-decoherence (Fig.~\ref{fig5} (f)), it is clearly seen that the position distribution along the $y$-direction spreads faster than that along the $x$-direction. The pattern along the $x$-direction reveals the classical binomial distribution (Fig.~\ref{fig4}).

\section{Correlations of the two-dimensioan AQW in the presence of decoherence}\label{IV}
When taking into the decoherence, we have discussed the anisotropic position distribution in 2D AQW. In this section, we quantitatively estimate the correlation between the walkers and the coin in the presence of decoherence. Exploiting the correlations reserved in the AQW is a very interesting problem, it will uncover how much correlations survive when the decoherence is introduced, and might have applications in the process of quantum information. Firstly, we study the time evolution of the correlations between the walker along the $x$-direction and the walker along the $y$-direction of 2D AQW. The classical mutual information is chosen to measure the classical correlation, and we use the measurement induced disturbance (MID) to qualify the quantum correlation~\cite{Ollivier_PRL2002, Luo_PRA2008, Xue_PRA2012}. Secondly, we qualify the quantum correlations stored between the walkers and the coin. Considering the correlations between the walkers along different directions and the coin, the anisotropic behaviors emerge. The definition of the classical mutual information $I_c(t)$ is presented below to measure the information shared by two walkers.
\begin{equation}
I_c(t)=\sum_x\sum_y P(x,y,t)\log_2(\frac{P(x,y,t)}{P(x,t)P(y,t)}).
\end{equation}
Here, $P(x,y,t)$ denotes the probability of the walkers occupying the position $(x,y)$ in the $x-y$ position space at time $t$. The marginal probability distribution $P(x,t)$ stands for the probability of the first walker occupying the position $x$ at time $t$, and the marginal probability distribution $P(y,t)$ represents the probability for the second walker occupying the position $y$ at time $t$. When two walkers are independent with each other, the shared information $I_c$ between them is zero.

For the quantum correlation, though it is well-estimated by quantum discord, it might be difficult to evaluate for the requirement of minimization over possible measurements~\cite{Ollivier_PRL2002, Luo_PRA2008}. Here, we use the MID $Q(\rho)$ to estimate the quantum correlation~\cite{Luo_PRA2008}, that is,
\begin{equation}
Q(\rho)=I(\rho)-I(\Pi\rho),
\end{equation}
with,
\begin{equation}
I(\rho)=S(\rho_1)+S(\rho_2)-S(\rho),
\end{equation}
where $S(\rho)=-\mathrm{Tr}(\rho\log_2\rho)$. The state $\Pi\rho$ satisfies $\Pi\rho=\sum_{j,k}\Pi_1^j\bigotimes\Pi_2^k\rho\Pi_1^j\bigotimes\Pi_2^k$. The projectors $\{\Pi_1^j\}$ and $\{\Pi_2^k\}$ represent the complete projective measurements which are performed on parties $1$ and $2$ of bipartite state $\rho$, respectively. We have the complete relations as $\rho_1=\sum_jp^j_1\Pi_1^j$ and $\rho_2=\sum_kp^k_2\Pi_2^k$. The reduced density matrix $\rho_1$ and $\rho_2$ are expressed as $\rho_1=\mathrm{Tr}_2\rho$ and $\rho_2=\mathrm{Tr}_1\rho$, respectively.
\begin{figure}[htbp]
\begin{center}
\includegraphics[width=0.5\textwidth]{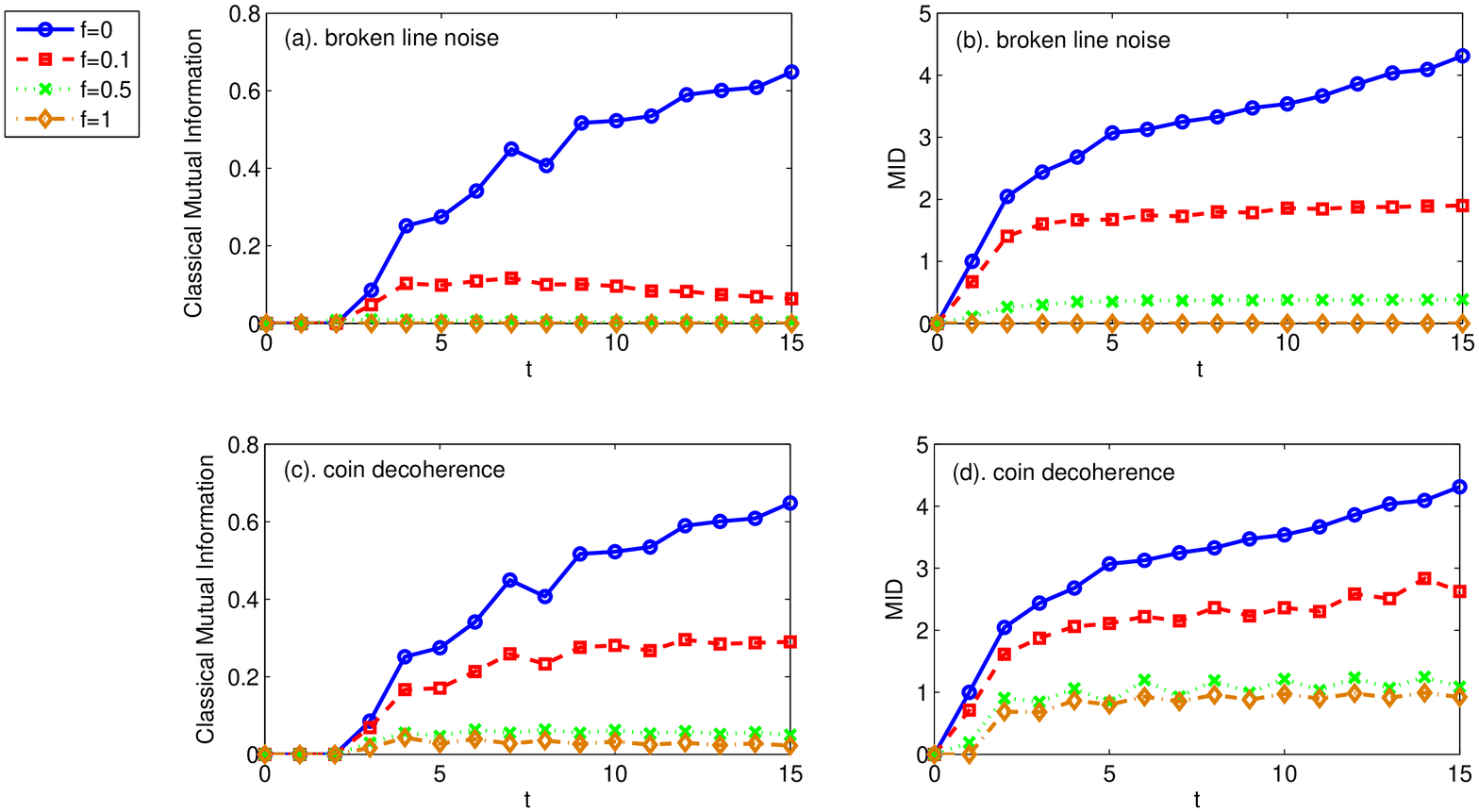}
\end{center}
\caption{\label{fig6} (Color online): The classical and quantum correlations between two walkers of 2D AQW in the presence of decoherence. (a) and (b), the broken line noise is introduced in the $x$-direction; (c) and (d), the coin-decoherence is applied on the coin before the walker travels along the $x$-direction. The classical correlation of two walkers is represented by classical mutual information, which are shown in (a) and (c). In (b) and (d), the lines for the quantum correlation represented by MID. Four different probabilities $f$ are chosen, blue circle, $f=0$; red rectangle, $f=0.1$; green cross, $f=0.5$; brown diamond, $f=1$. The initial state for the walkers and coin are, $|0\rangle_x|0\rangle_y\bigotimes(1/\sqrt{2}|R\rangle+i/\sqrt{2}|L\rangle)$.}
\end{figure}

The time evolutions of classical mutual information and MID between two walkers are shown in Fig.~\ref{fig6}. The figures above represent the 2D AQW affected by the broken line noise, and the figures below denote the AQW with the coin-decoherence. Four different probabilities $f$ are chosen for comparison. The 2D AQW without decoherence ($f=0$) is depicted in blue circle. The lines with red rectangle, green cross and brown diamond correspond to the 2D AQW with probability $f$ chosen as $f=0.1$, $f=0.5$ and $f=1$. When there is no decoherence in the AQW (blue circle in Fig.~\ref{fig6}), the correlations between the two walkers along different directions are strong. Under the influence of noises (red rectangle, green cross and brown diamond in Fig.~\ref{fig6}), the correlations between two walkers decrease. For the broken line noise model, when the probability $f$ approaches $1$ (brown diamond in Fig.~\ref{fig6} (a) and (b)), the classical and quantum correlations between two walkers are close to zero. As shown in Fig.~\ref{fig5}, the probability distribution of the AQW is trapped along the $x$-direction. Such localization of distribution reduces the correlation between two walkers to zero, and make the walkers travel along the $x$-direction and $y$-direction independently and share no information. When the coin-decoherence is introduced in the AQW, compared to the no-decoherent AQW, the correlation between two walkers reduces to a smaller value (Fig.~\ref{fig6} (c) and (d)). When the probability $f$ is close to $1$ (brown diamond in Fig.~\ref{fig6} (c) and (d)), the correlations between two walkers are not zero. As mentioned above, the coin-decoherence disturbs the 2D AQW weakly, and the motion in the $y$-direction exhibits the quantum behavior whatever the probability $f$ is taken. This weak decoherence affects motions of two walkers and keeps the certain correlations between them.

\begin{figure}[htbp]
\begin{center}
\includegraphics[width=0.5\textwidth]{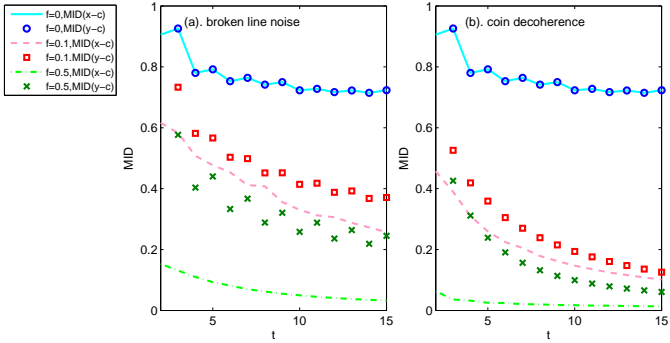}
\end{center}
\caption{\label{fig7} (Color online): The quantum correlations between walkers along different directions and the coin in the presence of decoherence. (a), the broken line noise is introduced in the $x$-direction; (b), the coin-decoherence is applied on the coin before the walker travels along the $x$-direction. The quantum correlation is represented by MID. Three different probabilities $f$ are chosen. For the correlation between the $x$-directional walker and the coin MID(x-c), light blue solid line, $f=0$; light red dashed line, $f=0.1$; light green dashed dotted line, $f=0.5$. For the correlation between the $y$-directional walker and the coin MID(y-c), dark blue circle, $f=0$; dark red rectangle, $f=0.1$; dark green cross, $f=0.5$. The initial state for the walkers and coin are, $|0\rangle_x|0\rangle_y\bigotimes(1/\sqrt{2}|R\rangle+i/\sqrt{2}|L\rangle)$.}
\end{figure}

Later, we qualify the correlations between the walkers along different directions and the coin. In our discussion, we study the quantum correlations estimated by MID between the $x$-directional walker and the coin MID(x-c), and between the $y$-directional walker and the coin MID(y-c), see Fig.~\ref{fig7}.
The time evolutions of quantum correlations MID(x-c) and MID(y-c) are drawn in Fig.~\ref{fig7}. Three different probabilities $f$ are chosen to illustrate the decoherence effect. For MID(x-c), the light blue solid, light red dashed, and light green dotted dashed lines correspond to the probability $f$ chosen as $f=0$, $f=0.1$ and $f=0.5$, respectively. For MID(y-c), the dark blue circle, dark red rectangle and dark green cross correspond to the probability $f$ chosen as $f=0$, $f=0.1$ and $f=0.5$, respectively. As shown in Fig.~\ref{fig7} (a) and (b), with the increase of probability $f$, both quantum correlations MID(x-c) and MID(y-c) decrease. For the decoherence appears only in one direction of 2D AQW, the anisotropic quantum correlations between two directional walkers and the coin emerge. In Fig.~\ref{fig7} (a) and (b), we find that, when there is no decoherence (blue solid line and blue circle in Fig.~\ref{fig7} (a) and (b)), the quantum correlations MID(x-c) and MID(y-c) have the same amplitudes with time. When the broken line noise or coin-decoherence is introduced into the walk, for the decoherence only emerges in the $x$-direction, the quantum correlation between the $x$-directional walker and the coin MID(x-c) is affected heavier than that between the $y$-directional walker and the coin MID(y-c). The remaining quantum correlation MID(y-c) is larger than the correlation MID(x-c). Referring to the variances of position distribution discussed above, when two kinds of decoherence are considered, the behavior along the $y$-direction exhibits more "quantumness" than that along the $x$-direction. In some sense, such anisotropic quantum correlations MID(x-c) and MID(y-c) correspond to the anisotropic position distribution patterns of 2D AQW aforementioned.

\section{Conclusions}\label{V}
In this paper, we have studied the dynamics of 2D AQW in the presence of decoherence. We present the analytic expressions for the first and second moments of the position distribution involving any kinds of decoherence. The emergence of quantum and classical behaviors in the AQW are discussed. Taking the broken line noise and coin-decoherence as examples of decoherence, we analyze the diffusion coefficients and the variances of position in the 2D AQW. We find that, when the broken line noise is applied on the system, the motions along $x$ and $y$-directions both change to exhibit the classical behaviors. When the broken probability $f$ approaches $1$, the walker along the $x$-direction is trapped, and the motion along $y$-direction displays quantum behaviors. In comparison, when the coin-decoherence is introduced into the walk, the coin is influenced by this weak decoherence, and only the dynamics along one direction is affected. In our study, we find that, the classical behaviors emerge in the motion along the $x$-direction, but the motion along the $y$-direction exhibits the quantum behaviors, whatever the strength of decoherence is taken.

In addition, we discuss the correlations between two walkers and coin in 2D AQW. Firstly, we employ the classical mutual information and MID to qualify the classical and quantum correlations between two walkers, respectively. We find that, with the appearance of the decoherence, the correlations between two walkers are smaller than those with no decoherence. For the broken line noise model, when the broken probability is close to $1$, the trap of the position distribution makes no correlations remain between two walkers. While for the case of coin-decoherence, such decoherence weakly disturbs the coherence evolution of the system, and the correlations between two walkers still exist no matter how strong the decoherence. Secondly, we discuss the quantum correlation between the walkers along different directions and the coin. For the decoherence is introduced along the $x$-direction, the anisotropic quantum correlations between two walkers and the coin emerge, which corresponds to the anisotropic position distribution of 2D AQW in the presence of decoherence.

Considering the 2D AQW has its advantages in designing of quantum search algorithms, and the decoherence from the surrounding environment is unavoidable, our study on the AQW incorporating the decoherence provides the theoretical basis for the development of quantum algorithms design and quantum information.

\section*{Acknowledgments}
We acknowledge the financial support from Young Teachers Academic Starting Plan No. 2015CX04046 of Beijing Institute of Technology.
{}

\end{document}